\documentstyle[11pt,newpasp,twoside,psfig]{article}
\def\edcomment#1{\iffalse\marginpar{\raggedright\sl#1\/}\else\relax\fi}
\reversemarginpar
\newcommand{\HI}{\protect\normalsize H\thinspace\protect\footnotesize
I\protect\normalsize} 

\newcommand{\etal}{{et~al.\, }}

\newcommand{\eg}{{e.g.\, }}         
\newcommand{\ie}{{i.e.\, }}         

\begin{document}
\title{An Overview of Optical Galaxy Searches and their Completeness}
 \author{Ren\'ee C. Kraan-Korteweg}
\affil{Depto.\ de Astronom\' \i a, Universidad de Guanajuato, 
Apdo.~Postal 144, Guanajuato GTO 36000, Mexico}

\begin{abstract}
Dust and stars in the Milky Way create a ``Zone of Avoidance'' (ZOA)
in the distribution of optically visible galaxies of about 25\% of the
sky. To reduce this gap, optical searches for partially obscured
galaxies have been performed. The status of these deep searches, in
particular their completeness as a function of foreground extinction
are discussed. Using existing sky surveys, over 50\,000 previously
unknown galaxies have been identified.  The surveys cover practically
the whole ZOA. It is shown that these surveys are complete for
galaxies with extinction-corrected diameters $D^o \ge 1.3$~arcmin to
extinction levels of $A_B \le 3\fm0$.  Incorporating these new data in
a whole-sky map of galaxies complete to $D^o \ge 1\farcm3$ finds the
ZOA reduced by a factor of about 2 to 2.5, respectively from
extinction levels of $A_B = 1\fm0$ to $A_B = 3\fm0$ (see Fig.~4).
\end{abstract}

\section{Historic Perspective of the Zone of Avoidance}
A first reference to the Zone of Avoidance (ZOA), or the ``Zone of few
Nebulae'' was made in 1878 by Proctor, based on the distribution of
nebulae in the ``General Catalogue of Nebulae'' by Sir John Herschel
(1864). This zone becomes considerably more prominent in the
distribution of nebulae presented by Charlier (1922) using data from
the ``New General Catalogue'' by Dreyer (1888, 1895). These data also reveal
first indications of large-scale structure: the nebulae display a very
clumpy distribution. Currently well-known galaxy clusters such as 
Virgo, Fornax, Perseus, Pisces and Coma are easily recognizable even
though Dreyer's catalog contains both Galactic and extragalactic objects.
It was not known then that the majority of the nebulae actually
are external stellar systems similar to the Milky Way. Even more
obvious in this distribution, though, is the absence of galaxies around
the Galactic Equator. As extinction was poorly known at that time, no
connection was made between the Milky Way and the ``Zone of few
Nebulae''.

A first definition of the ZOA was proposed by Shapley (1961) as the
region delimited by ``the isopleth of five galaxies per square degree
from the Lick and Harvard surveys'' (compared to a mean of 54
gal./sq.deg. found in unobscured regions by Shane \& Wirtanen
1967). This ``Zone of Avoidance'' used to be ``avoided'' by
astronomers interested in the extragalactic sky because of the lack of
data in that area of the sky and the inherent difficulties in
analyzing the few obscured galaxies known there.

Merging data from more recent galaxy catalogs, \ie the Uppsala General
Catalog UGC (Nilson 1973) for the north ($\delta \ge -2\fdg5$), the
ESO Uppsala Catalog (Lauberts 1982) for the south ($\delta \le
-17\fdg5$), and the Morphological Catalog of Galaxies MCG
(Vorontsov-Velyaminov \& Archipova 1963-74) for the strip inbetween
($-17\fdg5 < \delta < -2\fdg5$), a whole-sky galaxy catalog can be
defined. To homogenize the data determined by different groups from
different survey material, the following adjustments have to be
applied to the diameters: ${D = 1.15 \cdot D_{\rm UGC}, D = 0.96 \cdot
D_{\rm ESO}}$ and ${D = 1.29 \cdot D_{\rm MCG}}$ (Lahav
1987). According to Hudson \& Lynden-Bell (1991) this ``whole-sky''
catalog then is complete for galaxies larger than ${D} =1\farcm3$.

The distribution of these galaxies is displayed in Galactic
coordinates in Fig.~1 in an equal-area Aitoff projection
centered on the Galactic Bulge ($\ell = 0\deg, b = 0\deg$). The
galaxies are diameter-coded, so that structures relevant for the
dynamics in the local Universe stand out accordingly.  Figure~1
clearly displays the irregularity in the distribution of galaxies in
the nearby Universe such as the Local Supercluster visible as a great
circle (the Supergalactic Plane) centered on the Virgo cluster at
$\ell=284\deg, b=74\deg$, the Perseus-Pisces chain bending into the
ZOA at $\ell=95\deg$ and $\ell=165\deg$, the general overdensity in
the Cosmic Microwave Background dipole direction
($\ell=276\deg,b=30\deg$; Kogut \etal 1993) and the general galaxy
overdensity in the Great Attractor region (GA) centered on $\ell=320\deg,
b=0\deg$ (Kolatt, Dekel, \& Lahav 1995) with the Hydra ($270\deg,27\deg$), 
Antlia ($273\deg,19\deg$), Centaurus ($302\deg,22\deg$) and Pavo
($332\deg,-24\deg$) clusters.

\begin{figure}[ht]
\begin{center}
\hfil \psfig{figure=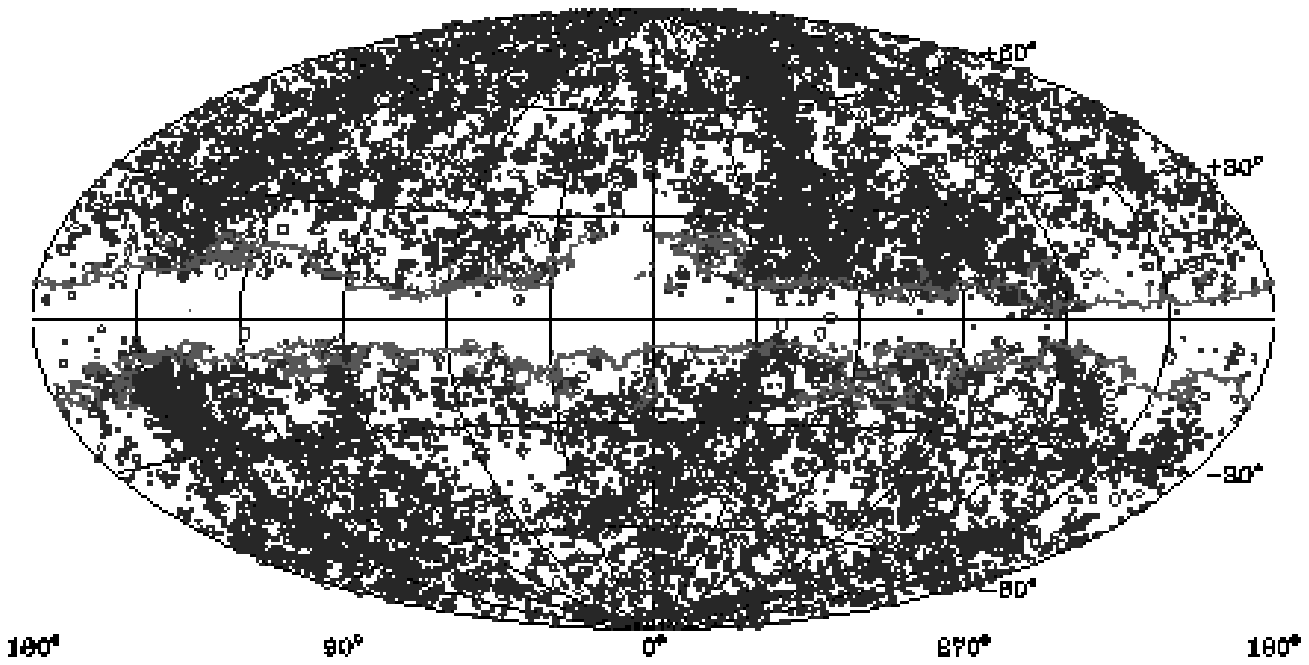,width=13cm} \hfil
\caption
{Aitoff equal-area projection in Galactic coordinates of galaxies with
${D}\ge1\farcm$3. The galaxies are diameter-coded: small circles
represent galaxies with $1{\farcm}3 \le {D} < 2\arcmin$, larger
circles $2\arcmin \le {D} < 3\arcmin$, and big circles ${D}
\ge 3\arcmin$. The contour marks absorption in the blue of ${A_B}
= 1\fm0$ as determined from the Schlegel \etal (1998) 
dust extinction
maps.  The displayed contour surrounds the area where the galaxy
distribution becomes incomplete (the ZOA) remarkably well.}
\end{center}
\end{figure}

Most conspicuous in this distribution is, however, the very broad,
nearly empty band of about 20$\deg$ width. As optical galaxy catalogs
are limited to the largest galaxies they become increasingly
incomplete close to the Galactic Equator where the dust thickens. This
diminishes the light emission of the galaxies and reduces their
visible extent. Such obscured galaxies are not included in the above
mentioned classic diameter- or magnitude-limited catalogs because they
appear small and faint -- even though they might be intrinsically
large and bright. A further complication is the growing number of
foreground stars close to the Galactic Plane which fully or partially
block the view of galaxy images.

Comparing this ``band of few galaxies'' with the currently available
100\micron\ dust extinction maps of the DIRBE experiment (Schlegel,
Finkbeiner, \& Davis 1998), we can see that the ZOA -- the area where
the galaxy counts become severely incomplete -- is described almost
perfectly by the absorption contour in the blue ${A_B}$ of $1\fm0$
(where ${A_B} = 4.14 \cdot E(B-V)$; Cardelli, Clayton, \& Mathis 1989). 
This contour
matches the by Shapley (1961) defined ZOA closely.

This wide gap, however, restricts a proper analysis of the dynamics in
our local Universe which requires whole-sky coverage, in particular
the determination of the peculiar velocity of the Local Group with
respect to the Cosmic Microwave Background and velocity flow fields
such as in the Great Attractor (GA) region (see Kraan-Korteweg \&
Lahav 2000, for a detailed review). The main uncertainty in, for
instance, the determination of the apex of the LG motion, as well as
the distance at which convergence is attained is dependent on the
width of the ZOA.  In the following, the achieved reduction of the ZOA
from systematic deep optical galaxy searches is presented.

\section{Status of Optical Galaxy Searches}

Using existing sky surveys such as the first and second generation
Palomar Observatory Sky Surveys POSS I and POSS II in the north, and
the ESO/SRC (United Kingdom Science Research Council) Southern Sky
Atlas, various groups have performed systematic deep searches for
``partially obscured'' galaxies. They catalogued galaxies down to
fainter magnitudes and smaller dimensions (${D} \ga 0\farcm1$) than
previous catalogs (e.g. ${D} \ge 1\farcm0$; Lauberts 1982). Here,
examination by eye remains the best technique. A separation of galaxy
and star images can as yet not be done on a viable basis below $|b|
\la 10\deg-15\deg$ by automated measuring machines and sophisticated
extraction algorithms, nor with the application of Artificial Neural
Networks (ANN). The latter was tested by Naim (1995) who used ANN to
identify galaxies with diameters above 25$\arcsec$ at low Galactic
latitudes ($b\sim 5\deg$). Galaxies could be identified using this
algorithm, and although an acceptable hit rate for galaxies of 80 --
96\% could be attained when ANN was trained on high latitude fields,
the false alarms were of equal order. Using low latitude fields as
training examples, the false alarms could be reduced to nearly zero
but then the hit rate was low ($\sim$ 30 - 40\%). Although the first
attempts of using ANN in the ZOA are encouraging they clearly need
further development.  Thus, although surveys by eye clearly are both
very trying and time consuming -- and maybe not as objective -- they
currently still provide the best technique to identify partially
obscured galaxies in crowded star fields.

Meanwhile, nearly the whole ZOA has been surveyed (see Fig.~2).  Over
50\,000 previously unknown galaxies could be discovered in this
way. The uncovered galaxies are not biased with respect to any
particular morphological type. A comparison of the survey regions with the
ZOA (Fig.~1) demonstrates that the systematic deep optical galaxy
searches cover practically the whole ZOA. Details and further
references about the surveys
and results on the uncovered galaxy distributions can be found for:\\
{A}:  the Perseus-Pisces Supercluster region in Pantoja et~al., 
these proceedings,\\
{B$_{1-3}$}: the northern Milky Way (Weinberger et~al., these
proceedings),\\
{C$_{1}$}: the Puppis region (Saito \etal 1990, 1991),\\
{C$_{2-3}$}: the Sagittarius and Aquila region (Roman \& Saito, 
these proceedings),\\
{D$_{1}$}: the Puppis/Hydra region (Salem \& Kraan-Korteweg, in prep.),\\
{D$_{2}$}: the Hydra/Antlia region (Kraan-Korteweg 2000),\\
{D$_{3-4}$}: the Crux and Great Attractor region (Woudt et~al., these proceedings),\\
{D$_5$}: the Scorpius region (Fairall \& Kraan-Korteweg, these proceedings),\\
{E}: the Ophiuchus region (Wakamatsu et~al., these proceedings),\\
{F}: the northern GP/SGP crossing (Hau \etal 1995).

\begin{figure}[t]
\begin{center}
\hfil \psfig{figure=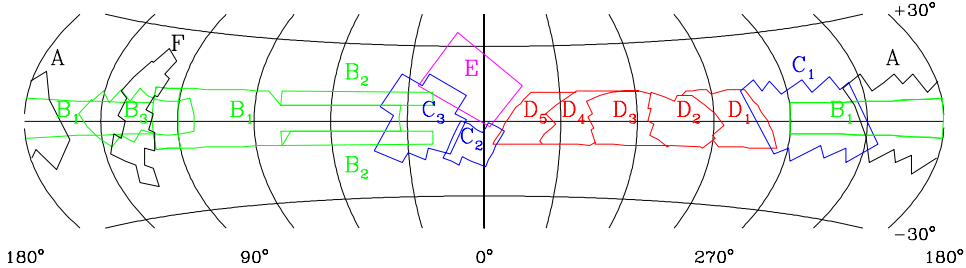,width=13cm} \hfil
\caption
{An overview of the different optical galaxy surveys in the ZOA centered
on the Galaxy. The labels identifying the search areas are explained 
in the text. Note that the surveyed regions cover the entire ZOA as
defined by the foreground extinction level of ${A_B} = 1\fm0$
displayed in Fig.~1.}
\end{center}
\end{figure}

\section{Completeness of Optical Galaxy Searches}
Most of these searches have quite similar characteristics.
The distributions reveal that galaxies can easily be traced through
obscuration layers of 3 magnitudes (see Fig.~2 in Fairall \& 
Kraan-Korteweg, these proceedings), thereby narrowing the ZOA
considerably. A few galaxies are still recognizable up to extinction
levels of $A_B = 5\fm0$ and a handful of very small galaxy
candidates have been found at even higher extinction levels. Overall, the
mean number density follows the dust distribution remarkably well at
low Galactic latitudes. The contour level of $A_B = 5\fm0$, for
instance, is nearly indistinguishable from the galaxy density contour
at 0.5 galaxies per square degree (Kraan-Korteweg 1992; Woudt 1998).

The galaxies detected in these searches are quite small ($<D> =
0\farcm4$) on average. So the question arises whether these new
galaxies are relevant at all to our understanding of the dynamics in
the local Unverse. Analyzing the galaxy density as a function of
galaxy size, magnitude and/or morphology in combination with the
foreground extinction has led to the identification of various
distinct structures and their approximate distances.  However, a
detailed understanding of the completeness as a function of the
foreground extinction in such searches is essential for the
disentanglement of large-scale structure and absorption.  Therefore,
Kraan-Korteweg (2000) and Woudt (1998) studied the apparent diameter
distribution as a function of the extinction $E(B-V)$ (Schlegel \etal
1998) as well as the location of the flattening in the slope of the
cumulative diameter curves $(\log D) - (\log N)$ for various
extinction intervals for the galaxies in their ZOA galaxy catalogs
(D$_{2-4}$ in Fig.~2). Inspecting for instance the top panels of
Fig.~3, it can be seen that the deep optical survey is complete to an
apparent diameter of ${D} = 14\arcsec$ ($\log D = 1.15$)
for extinction levels less than $E(B-V) = 0.72$, i.e. $A_{B} = 3\fm0$.

\begin{figure}[t]
\begin{center}
\hfil \psfig{figure=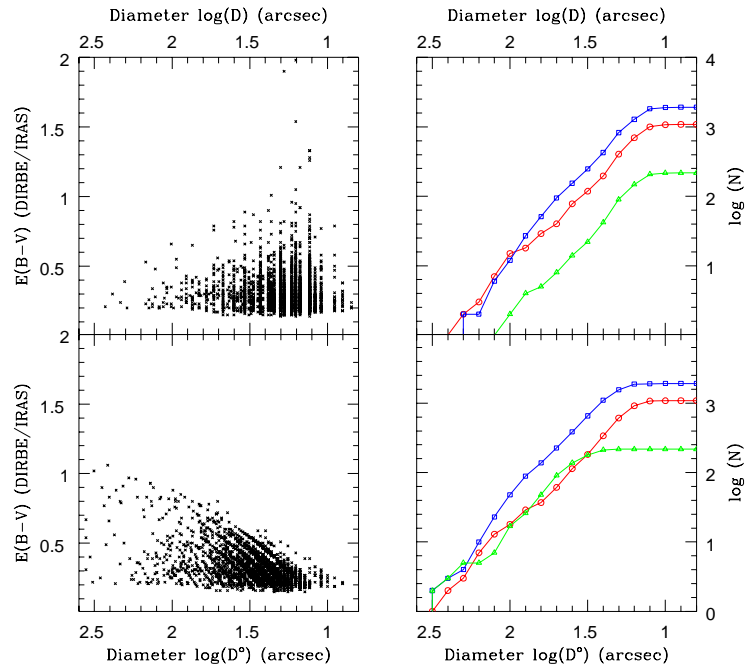,width=10cm} 
\caption
{Observed (top panel) and extinction-corrected (bottom) diameters of
ZOA galaxies. Left panels: diameters as a function of the foreground
extinction $E(B-V)$. Right panels: the cumulative diameter
distributions for three intervals of Galactic foreground extinction
(open circles: $A_B \le 1^{\rm m}$; squares: $1^{\rm m} < A_B \le
2^{\rm m}$; triangles: $2^{\rm m} < A_B \le 3^{\rm m}$).
}
\end{center}
\end{figure}

How about the intrinsic diameters, \ie the diameters galaxies would
have if they were unobscured? A spiral galaxy seen through an
extinction of $A_{B} = 1\fm0$ will, for example, be reduced to $\sim
80\%$ of its unobscured size.  Only $\sim 22\%$ of a (spiral) galaxy's
original dimension is seen when it is observed through $A_{B} =
3\fm0$.  In 1990, Cameron investigated these obscuration effects by
artificially absorbing the intensity profiles of unobscured galaxies.
He found a diameter reduction due to the extinction $A_B$ of $f =
10^{0.13 {A_B}^{1.3}}$ for ellipticals/lenticulars, and of $f =
10^{0.10 {A_B}^{1.7}}$ for spirals.

When applying these corrections to the observed (reduced) diameters,
it becomes clear that \eg an obscured spiral or an elliptical galaxy
at the {\it apparent} completeness limit of ${D} = 14\arcsec$ seen
through an extinction layer of \eg at $A_{B} = 3\fm0$ has an {\it
intrinsic} diameter of ${D^o} \sim 60\arcsec$, respectively ${D^o}
\sim 50\arcsec$. At extinction levels higher than $A_{B} = 3\fm0$, an
elliptical galaxy with $D^o = 60\arcsec$ would appear smaller than the
completeness limit $D = 14\arcsec$ and might have gone unnoticed.
Optical galaxy catalogs should therefore be {\it complete} to $D^o \ge
60\arcsec$ for galaxies of all morphological types down to extinction
levels of $A_{B} \le 3\fm0$ with the possible exception of extremely
low-surface brightness galaxies. Only intrinsically very large and
bright galaxies -- particularly galaxies with high surface brightness
-- will be recovered in deeper extinction layers. This completeness
limit could be confirmed by independently analyzing the diameter
vs. extinction and the cumulative diagrams of extinction-corrected
diameters (see bottom panels of Fig.~3).

It can thus be presumed that all the similarly performed optical
galaxy searches in the ZOA are complete for galaxies with
extinction-corrected diameters ${D^o} \ge 1\farcm0$ to extinction
levels of $A_{B} \le 3\fm0$. As the completeness limit of the optical
searches lies well below the completeness limit $D = 1\farcm3$ of the
ESO, UGC and MGC catalogs one can then supplement these catalogs
with the galaxies from optical ZOA galaxy searches that have 
${D^o} \ge 1\farcm3$ and $A_{B} \le 3\fm0$.

This has been done in Fig.~4 which shows an improved whole-sky galaxy
distribution with a reduced ZOA. In this Aitoff projection, all the
UGC, ESO, MGC galaxies that have {\it extinction-corrected} diameters
${D^o} \ge 1\farcm3$ are plotted (remember that galaxies adjacent to
the optical galaxy search regions are also affected by absorption
though to a lesser extent: $A_{B} \la 1\fm0$), next to all the
galaxies from the various optical surveys with ${D^o} = 1\farcm3$ and
$A_{B} \le 3\fm0$ for which positions and diameters were
available. The regions for which these data are not yet available are
outlined in Fig.~4. As some searches were performed on older generation
POSS I plates, which are less deep compared to the second generation
POSS II and ESO/SERC plates, an additional correction was applied to
those diameters, \ie the same correction as for the UGC galaxies which
also are based on POSS I survey material (${D_{25} = 1.15 \cdot D_{\rm
POSS I}}$).

\begin{figure}[ht]
\begin{center}
\hfil \psfig{figure=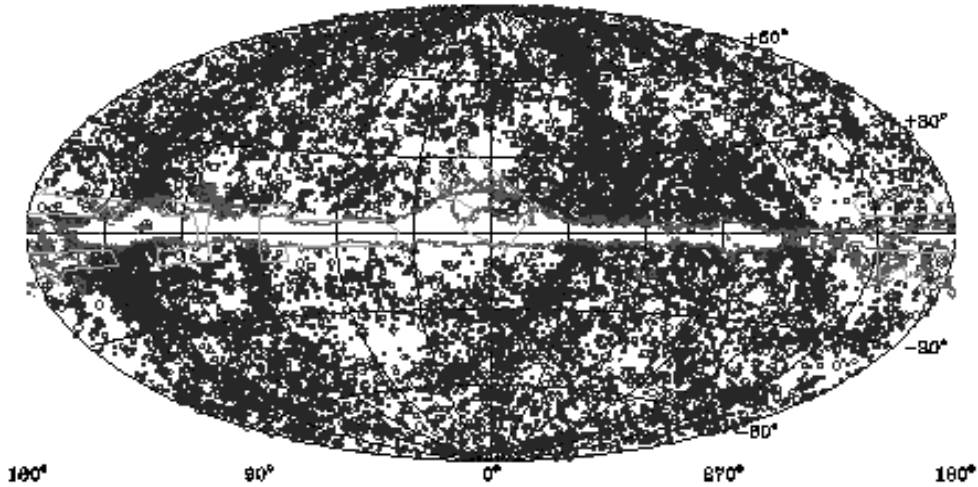,width=13cm} \hfil
\caption
{Aitoff equal-area distribution of ESO, UGC, MCG galaxies with 
extinction-corrected diameters ${D^o} \ge 1\farcm3$, including 
galaxies identified in the optical ZOA galaxy searches for 
extinction-levels of $A_{B} \le 3\fm0$ (contour). The diameters are
coded as in Fig.~1. With the exception of the areas for
which either the positions of the galaxies or their diameters are
not yet available (demarcated areas), the ZOA could be reduced
considerably compared to Fig.~1.}
\label{aitc}
\end{center}
\end{figure}

A comparison of Fig.~1 with Fig.~4 demonstrates convincingly how the
deep optical galaxy searches realize a considerable reduction of the
ZOA: we can now trace the large-scale structures in the nearby
Universe to extinction levels of $A_{B} = 3\fm0$. Inspection of Fig.~4
reveals that the galaxy density enhancement in the GA region is even
more pronounced (see for instance Woudt et~al., these proceedings, for
details on the uncovered rich cluster A3627 in the GA region) and a
connection of the Perseus-Pisces chain across the Milky Way at
$\ell=165\deg$ is more likely. Hence, these supplemented whole-sky maps
certainly should improve our understanding of the velocity flow fields
and the total gravitational attraction on the Local Group.

\section{Conclusion}

In the last decade, enormous progress has been made in unveiling
galaxies behind the Milky Way. At optical wavebands, the entire ZOA
has been systematically surveyed.  These surveys are complete for
galaxies larger than $D^o = 1\farcm3$ (corrected for absorption) down
to extinction levels of ${A_B} = 3\fm0$. Combining these data with
previous ``whole-sky'' maps reduces the ``optical ZOA'' by a factor of
about 2 - 2.5, which allows an improved understanding of the velocity
flow fields and the total gravitational attraction on the Local
Group. Various previously unknown structures in the nearby Universe
could be mapped in this way.

A difficult task is still awaiting us, \ie to obtain a detailed
understanding of the selection effects in the various searches in
which different groups identified galaxies from partly different plate
material. Quantifying the selection effects is crucial for any optimal
reconstruction method and important for quantitative
cosmography. Moreover, we need a better understanding of the effects
of obscuration on the observed properties of galaxies, i.e. the
Cameron corrections, in addition to an accurate high-resolution,
well-calibrated map of the Galactic extinction.

The remaining optical ZOA might yet hide further dynamically important
galaxy densities. Here, systematic surveys at other wavebands such as
\HI, near and far infrared, and X-ray become more efficient. The
success and status of these approaches are discussed in various
chapters in these proceedings and reviewed in Kraan-Korteweg \& Lahav
2000.

\end{document}